# A Distributed Sensor Data Search Platform for Internet of Things environments

Luiz H. Nunes[1,2], Júlio C. Estrella[1], Luis H. V. Nakamura[1,2], Rafael M. de O. Libardi[1],
Carlos H. G. Ferreira[1], Liuri L. R. Jorge[1], Charith Perera[3], Stephan Reiff-Marganiec[4]

[1] University of São Paulo, Institute of Mathematics and Computer Science (ICMC), São Carlos - SP, Brazil.
[2] Instituto Federal de São Paulo (IFSP), Matão-SP, Brazil.
[3] Faculty of Maths, Computing and Technology, The Open University, Walton Hall,
Milton Keynes, MK7 6AA-UK
[4] University of Leicester, University Road, Leicester, LE1 7RH - UK.
{lhnunes, , jcezar, nakamura, mira, chgferreira, liuri.jorge }@icmc.usp.br
charith.perera@ieee.org
srm13@le.ac.uk

## Abstract

Recently, the number of devices has grown increasingly and it is hoped that, between 2015 and 2016, 20 billion devices will be connected to the Internet and this market will move around 91.5 billion dollars. The Internet of Things (IoT) is composed of small sensors and actuators embedded in objects with Internet access and will play a key role in solving many challenges faced in today's society. However, the real capacity of IoT concepts is constrained as the current sensor networks usually do not exchange information with other sources. In this paper, we propose the Visual Search for Internet of Things (ViSIoT) platform to help technical and non-technical users to discover and use sensors as a service for different application purposes. As a proof of concept, a real case study is used to generate weather condition reports to support rheumatism patients. This case study was executed in a working prototype and a performance evaluation is presented.

_______________________________________________

## 1. Introduction

In recent years mobile devices, such as smartphones and tablets, had their costs reduced and their processing capability increased. In this way, the number of Internet connected devices exceeded the number of the world's population (about 6.3 billion people), between 2008 and 2009 was marked the beginning of the Internet of Things (IoT) [Research, 2011]. Recently, the number of devices has grown rapidly and it is hoped that between 2015 and 2016 about 20 billion devices will be connected to the Internet and having a market value of around 91.5 billion dollars [Evans, 2011].

The IoT is composed of small sensors and actuators embedded in objects such as electronic devices (e.g. smartphones or tablets), clothes, alarm systems, cars, domestic appliances and industrial machines, which are capable of interacting with each other through protocols using the Internet

[Bari et al., 2013], [Parwekar, 2011]. The mix of embedded devices with sensor networks allows to connect the real world with cyberspace and enables the deployment of new kinds of services and applications [Fan and Zhou, 2011], [Parwekar, 2011]. Environmental monitoring, smart homes and smart buildings are examples of recent applications of IoT concepts [Sanchez López et al., 2012].

It is expected that IoT will play a key role in solving many challenges faced in today's society [Koreshoff et al., 2013]. However, current applications just aim to solve problems in specific environments, where a private sensor network is set up and helps to build a closed information flow. The real capacity of IoT concepts is constrained, because these private sensor networks do not exchange information with others sources or users [Wirtz and Wehrle, 2013].

The collaboration between private sensors network can help to develop solutions for different problems as they obtain large amounts of reusable data for different purposes. Such data sets present opportunities to develop unprecedented services. For example, sensors can be used to efficiently manage the power consumption of a region or recognize patterns that predict and detect natural disasters [Zhang et al., 2013].

Despite the proliferation of cloud computing models and infrastructure, there is no simple way to manage environments to explore the features offered by the different devices that comprises IoT [Soldatos et al., 2012]. Due to their heterogeneity, costs and complexity these environment are usually represented through simulations or in very small scale sensor networks. OpenIoT [1], GSN [2] and Irisnet [3] are middleware that integrate different sensors and enable users to access the gathered data. Usually, these middleware use the concept of virtual-sensor to abstract the physical properties of one or more sensors and to handle with the data flow [Vermesan, 2014].

Unlike the solutions proposed so far which emphasize the integration of different sensors for data analysis, in this work we present the Visual Search for Internet of Things (ViSIoT) platform. The main concern of ViSIoT is to help technical and non-technical users to discover and use sensors as a service for different application purposes. As a proof of concept, a working prototype and a performance evaluation are presented. Also, ViSIoT will be used to generate a report about the weather conditions in Europe (EU) and North America (NA) between February 7 and 9, 2015 looking to advise rheumatism patients. The main contribution of ViSIoT can be summarized as: 1) feeds the existing Sensing as a Service solutions with virtual sensors distributed in a world-scale available in public cloud repositories; 2) provides the integration of multiple repositories and solutions; and 3) abstracts the IoT environment complexity providing a user interface to set up and deploy desired sensors into the specified devices.

_______________________________________________

1.   OpenIoT project - http://openiot.eu
2.   GSN project - http://sourceforge.net/projects/gsn/
3.   Irisnet project - http://www.intel-iris.net/

The rest of this paper is organized as follows: Section 2 presents a literature review. Section 3 describes the proposed platform and how it works. Section 4 describes our prototype development. Section 5 describes our motivational example and how ViSIoT applies to it. Section 7 presents the case study and ViSIoT performance results. Finally, the conclusions and directions for future work are presented in Section 8.

## 2. BACKGROUND AND RELATED WORKS

The combination of sensor networks and cloud computing models can offer data or events from these sensors as a service over the Internet. Several middleware systems for IoT seek to provide a layer between the infrastructure and applications to abstract the technological details allowing users to focus on developing applications for IoT [Chaqfeh and Mohamed, 2012]. The following examples show how existing IoT middleware solutions provide sensor search functionality and their data.

Linked sensor middleware (LSM) [Le-Phuoc et al., ], is a platform that joins the sensor data with the Semantic Web in an unified model. However, all data access and sensor searching must be done using SPARQL and a web interface, which is not user-friendly to non-technical users. Similar to LSM, Microsoft SensorMap based on the Microsoft SensorWeb platform uses a map and keywords to provide access to sensor data [Nath et al., 2007]. Global Sensor Networks (GSN), is a platform to integrate heterogeneous sensing technologies through a peer-to-peer model. It uses virtual sensors, which are offered as a service and abstracts the data collection process and processing of one or more sensors. The sensor identification and discovery uses keywords and the selection of the sensors is possible through a list of checkboxes on a web interface [Aberer and Hauswirth, 2006]. The OpenIoT platform uses the GSN platform to provide dynamic searching and data access using ontologies and semantic structures [Soldatos et al., 2012].

We now briefly describe some of the work done in sensing as a service mechanisms. Zhang et al. (2013) propose a platform to provide a unified view of data and workflow to maximize the sharing and utility of available sensor data sources, data, and data processing tools, to enable greater sensor data services. It shows a case study of their architecture that uses 60 firefly devices deployed over the Building 23 at CMUSV. Casola et al. (2013) present a Cloud infrastructure to ensure the SLA in sensor network as a service, which aggregates different network providers offering access to their private sensor networks to clients having specific requirements. They validated their approach using a testbed composed of eight sensors grouped in two networks with four sensors each. Mayer et al. (2012) show a prototype implementation of a Web-based infrastructure for smart devices to offer scalability, location-awareness, self-management, and user-friendliness, which were validated through the simulation of six hundred sensors of different types (e.g., temperature, electricity consumption, ambient light).

Searching and selection mechanisms also gained much attention in the Sensing as a Service field. Elahi et al. (2009) used prediction models to rank sensors according to their matching probability of a content-based sensor search. They used two real-world data sets totaling two hundred and fifty sensors to show the performance improvement of their search engine compared to a baseline method. Ostermaier et al. (2010) present the Dyser search engine for the Web of Things, which also uses prediction models to rank the available sensors. Dyser performance was evaluated using a real-world data set composed by 385 sensors over a period of five months. Calbimonte et al. (2011) presented an ontology-based framework for querying sensor data considering meta-data and mappings to underlying data sources. A federated sensor network environment with approximately one thousand and three hundred sensor was used as testbed. Perera et al. (2014) introduce a context-aware sensor search, selection, and ranking model, called CASSARAM to address the challenge of efficiently selecting a subset of relevant sensors out of a large set of sensors. Their testbed is composed of more than 100,000 sensors descriptions captured from real datasets.

Table I. Summary of the number of sensors provided by each solution. (Adapted from Perera et. al. (2014) and Farooq and Kunz (2014))

| Approach | Type of sensors | Number of nodes |
|---|---|---|
| Elahi et al. (2009) | Real Datasets | 250 |
| Ostermaier et al. (2010) | Real Datasets | 385 |
| Calbimonte et al (2011) | Real | 1300 |
| Mayer et al. (2012) | Simulation | 600 |
| Casola et al. (2013) | Real | 8 |
| Zhang et al. (2013) | Real | 60 |
| Perera et al. (2014) | Sensors descriptions | 100,000 |
| WISEBED | Real | 711 |
| SensLAB | Real | 1000 |

Although these papers handle different problems for sensing as a service they perform experiments with either a few real nodes or simulated nodes using real old datasets          due to the high cost to set up a sensor network environment and lack of interoperability. Farooq and Kunz (2014) present a survey with the available public testbeds projects for sensors networks where the WISEBED [Chatzigiannakis et al., 2010] and SensLAB [des Roziers et al., 2011] projects offer larger numbers of real sensor nodes with seven hundred and eleven and one thousand sensors respectively.

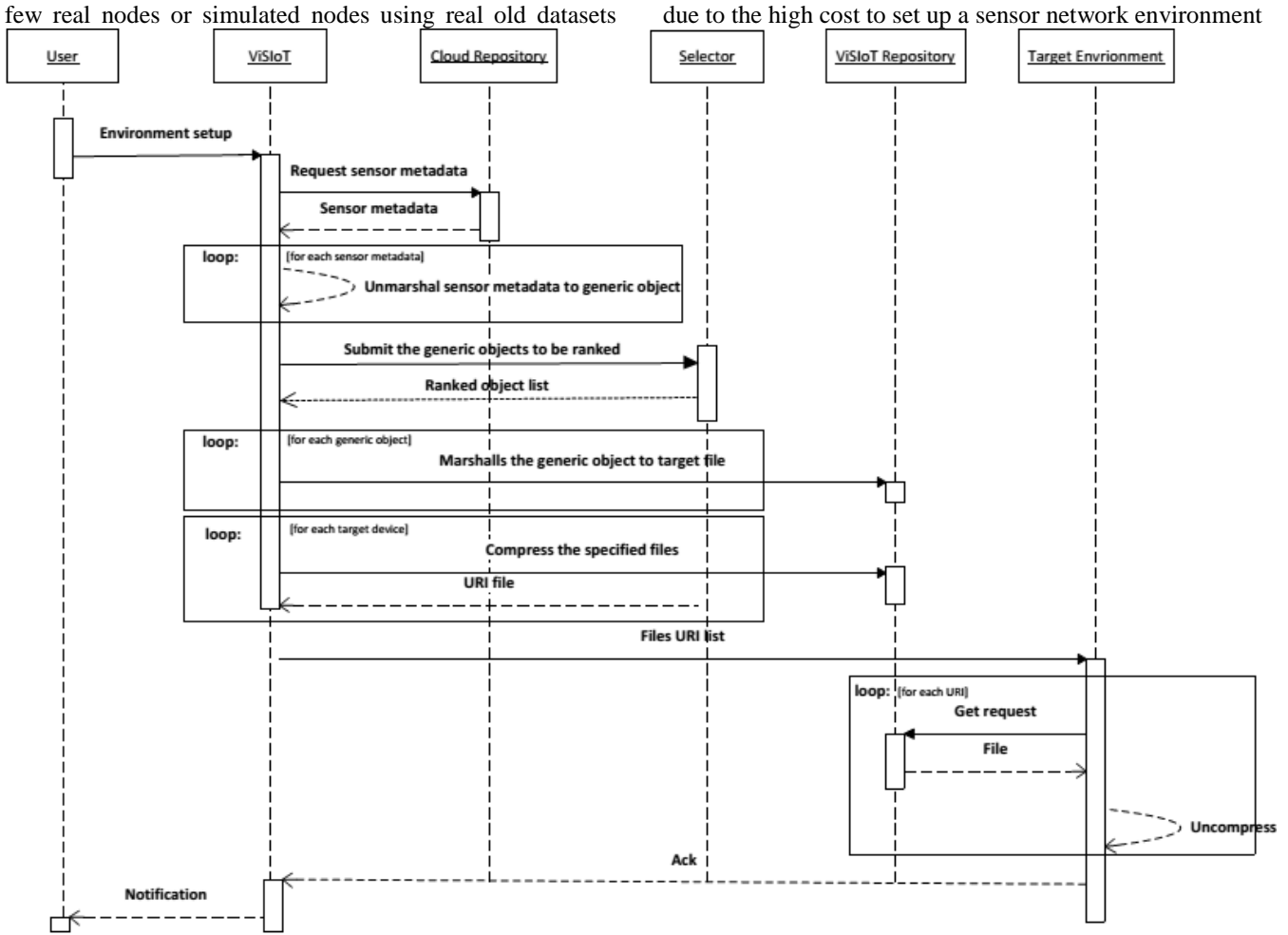

*Figure 1. ViSIoT Sequence Diagram*

In summary, it is possible to define three major problems to develop sensing as a services mechanisms. The first problem is the lack of a testbed with a huge number of sensors available for researchers, like described in Table I. The second problem is that most of the testbeds provided just cover a specific small area. Finally, the available sensor networks used as testbeds do not provide any interface to be re-used by another sensing as service solutions. ViSIoT differs from the works discussed in this section because it provides a platform to access a huge number of sensors distributed around the world, which gets real time data and can be accessed for any sensing as a service solution.

## 3. ViSIoT Architecture

The architecture of our proposed tool is based on a client-server architecture for distributed experiments in ser-vices oriented systems described in Nunes et al. (2015) and Nunes et al. (2014). These studies are used as a baseline of our work as they are able to successfully setup distributed environments based on user requirements.

The ViSIoT architecture provides access to a set of public virtual-sensors available as a service which can feed any type of sensing application or middleware for sensing as a service. Also ViSIoT can abstract the environment complexity using a client application to deploy the sensors into multiples target devices.

Figure 1 shows how ViSIoT architecture works. The sequence of steps performed by ViSIoT can be summarized as:

1)   The user sets the desired number of sensors, their location and arrangement in the target environment;

2)   ViSIoT performs requests to the selected cloud sensor repository to get the available sensors;

3)   ViSIoT unmarshals the response message into a generic sensor object. This generic object has specific information

about the sensors such as coordinates, type and how to access them;

4) ViSIoT ranks the generic objects to get the best available objects

5) ViSIoT marshals the generic sensor object into the target format specification. The marshaled object is stored into a local repository.

6) ViSIoT compress into a single file the set of files to be deployed into a target device

7) ViSIoT receives the URI of the marshaled file for external access;

8) ViSIoT send the selected URI list to the target device;

9) The target devices will download the file from the repository and uncompress.

## 4. ViSIoT Prototype

The prototyping platform was implemented as a proof of concept of our architecture using the OpenWeatherMap1 as a cloud sensor repository and the GSN as the target system. An user interface is also incorporated into the original architecture to conduct the environment setup. Figure 2 describes the prototype modules organization:

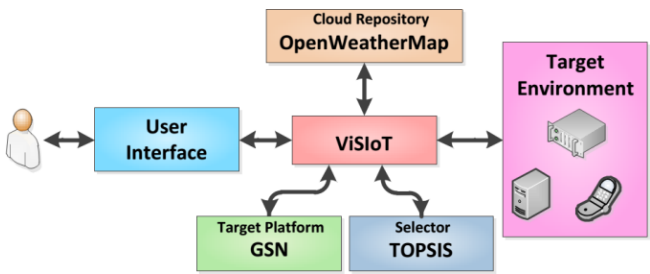

*Figure 2. ViSIoT prototype*

### 4.1 Cloud Repository

The Cloud Repository module represents any sensor repository that is available in the cloud. It must provide relevant information about the sensors and their state such as their coordinates, battery level and price. The data format provided by these repositories must be handled by the core of application.

The OpenWeatherMap[4] API is used as cloud sensor repository of our prototype. OpenWeatherMap weather service uses the OWM platform to collect, process, and distribute information about the world through easy tools and APIs. It has more than 40,000 weather stations around the world, which are installed in airports, large cities or even offered by fans and weather enthusiasts. Data gathered by OpenWeather is available in JSON, XML, or HTML format. Listing 1 corresponds to the response message for the following call

http://api.openweathermap.org/data/2.5/weather?lat=35&lon=139. The OpenWeatherMap response message can be unmarshaled into one or more virtual-sensors containing information like temperature, humidity and pressure.

LISTING 1. OPENWEATHERMAP JSON EXAMPLE

{"coord":{"lon":139,"lat":35},
"sys":{"country":"JP","sunrise":1369769524,
"sunset":1369821049},
"weather":[{"id":804,"main":"clouds",
"description":"overcast clouds","icon":"04n"}],
"main":{"temp":289.5,"humidity":89,"pressure":1013,
"temp_min":287.04,"temp_max":292.04},
"wind":{"speed":7.31,"deg":187.002},
"rain":{"3h":0},
"clouds":{"all":92},
"dt":1369824698,
"id":1851632,
"name":"Shuzenji",
"cod":200}

### 4.2 Target Platform

The target platform module represents the system that will handle the data in the target environment. Analogous to Section 4.1, the core of the application must provide the mechanisms to convert the sensor from the cloud repository to a sensor that can be used by the system.

LISTING 2. GSN VIRTUAL-SENSOR DESCRIPTION

```xml
<?xml version="1.0" encoding="UTF-8" standalone="no"?>
<virtual-sensor name="Tartu588335" priority="10">
 <processing-class>
  <class-name>gsn.vsensor.BridgeVirtualSensor</class-name>
  <init-params/>
  <output-structure>
   <field name="city" type="varchar(255)"/>
   <field name="country" type="varchar(255)"/>
   <field name="base" type="varchar(255)"/>
   <field name="temp" type="double"/>
   <field name="sea_level" type="double"/>
   <field name="pressure" type="double"/>
   <field name="humidity" type="double"/>
  </output-structure>
 </processing-class>
 <description/>
 <addressing>
  <predicate key="geographical">Tartu 588335</predicate>
  <predicate key="LATITUDE">26.72509</predicate>
  <predicate key="LONGITUDE">58.380619</predicate>
 </addressing>
 <storage history-size="168h"/>
 <streams>
  <stream name="stream1">
   <source alias="source1" sampling-rate="1"
    storage-size="168h">
    <address wrapper="openweathermap">
     <predicate key="url">http://api.openweathermap.org/
      data/2.5/weather?id=588335</predicate>
     <predicate key="type">humidity</predicate>
    </address>
    <query>Select city, country, base, sea_level, temp,
     humidity, pressure from wrapper</query>
   </source>
   <query>Select * from source1</query>
  </stream>
 </streams>
</virtual-sensor>
```

Global Sensor Network (GSN) is a middleware which supports the deployment, integration and discovery of a wide range of sensor network technologies through virtual which enables the user to specify XML-based deployment

---

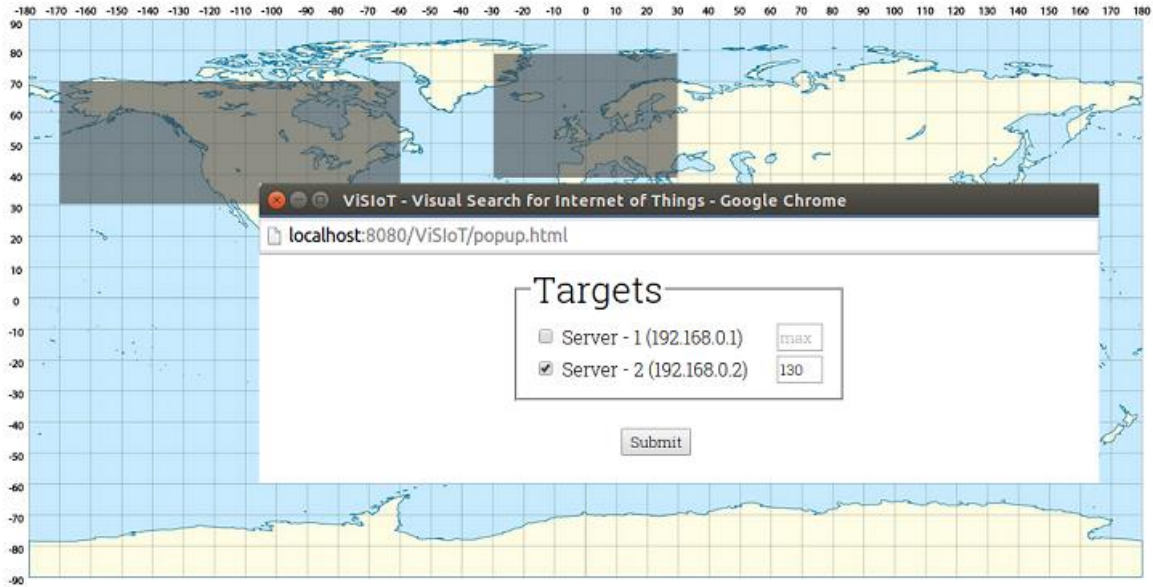

*Figure 3. ViSIoT User Interface*

sensors. A virtual sensor is a powerful sensor abstraction descriptors to integrate local and remote sensor data sources using SQL queries [Aberer and Hauswirth, 2006].

Listing 2 shows an example which defines a virtual sensor that reads the humidity from a specified city and returns the value to the user. This XML contains the information of a specific sensor such as coordinates, type of sensor, how to access it and the class that will process the virtual sensor file.

## 4.3 User Interface

The user interface provides an easy way to select several virtual sensors and configures the desired IoT environ-ment. It uses a world map to enable the user to select the desired region as shown in Figure 3 represented by the dark squares. When a region is selected, a pop-up window expands automatically and allows the user to select the deployed target system for that region and the number of desired sensors. It is also possible to limit the number of sensor that are deployed in the target devices.

## 4.4 Selector

The Selector module aims to rank the sensors available in a region. The ranking process should consider the sensors attributes to establish which are the best sen-sors. The ViSIoT prototype uses the Technique for the Order of Prioritisation by Similarity to Ideal Solution (TOPSIS) [Opricovic and Tzeng, 2004] to establish the most efficient trade-off between the attributes of a set of sensors. TOPSIS method has been applied in several areas such as Supply Chain Management and Logistics, Design, Engineering and Manufacturing Systems, Busi-ness and Marketing Management, Health, Safety and Environment Management, Human Resources Manage-ment, Energy Management, Chemical Engineering and Water Resources Management [Behzadian et al., 2012].

1) Normalize the analysis matrix Q to Q' according to the Equation 1:

$$q'_{ij} = \frac{q_{ij}}{\sqrt{\sum_{i=1}^{N}(q_{ij})^2}} \qquad (1)$$

where N represents the number of criteria in the evaluation matrix.

2) Determine the positive ideal points ($p_{+j}$) and the negative ideal points ($p_{-j}$) of all criteria using the analysis matrix. For a maximization criterion, the positive ideal and the negative ideal points can be calculated using Equations 2 and 3, respectively:

$$p_{+j} = \max_i(q'_{ij}) \qquad (2)$$

$$p_{-j} = \min_i(q'_{ij}) \qquad (3)$$

3) Compute the distances to the positive ideal solution and ($si_+$) and the negative ideal solution ($si_-$). The distance of each option q0 to the ideal solution $p_{+j}$ and the ideal negative solution $p_{-j}$ j is given by Equations 4 and 5:

$$s_{i+} = \sqrt{\sum_{j=1}^{n}(q'_{ij} - p_{+j})^2} \qquad and \qquad (4)$$

$$s_{i-} = \sqrt{\sum_{j=1}^{n}(q'_{ij} - p_{-j})^2} \qquad (5)$$

4) Calculate the relative closeness to the ideal solution. The relative closeness of $\mathbf{q}$ to $p_{-j}$ and $p_{+j}$ represented by ($ci_+$) can be calculated according to Equation 6.

$$c_{i+} = \frac{s_{i-}}{s_{i+} - si-} \qquad (6)$$

5) Sort options $q_i$ in increasing order according to the relative closeness to $ci_+$.

## 4.5 ViSIoT Core

The core structure is responsible to integrate all other modules and configure the desired environment. First, ViSIoT core communicates with the interface and gets the users sensors configuration. ViSIoT core connects in the example to the OpenWeatherMap and makes a set of requests to get the desired sensor group. The sensors are randomly chosen based on the received query result. The received messages follow the same structure as defined in Listing 1 and are unmarshaled to the defined generic object structure.

All desired sensors are unmarshaled to generic objects and then marshaled to the GSN virtual sensor format as described in Listing 2. After this step, the desired group of sensor for each client are compressed into a file, which is available for download. Finally, a RESTful message with a set of compressed file URIs is submitted to the target system. In addition, ViSIoT is structured to support several kinds of cloud sensors repositories and middleware for sensing as a service. The four main classes of ViSIoT are:

**Repository:** is responsible to get the sensors into the cloud repository. The RequestToRepository module man-ages the sensors that will be available for the experiments. It must have the send() and unmarshal() operation. The send() operation request the sensors from the cloud repository and returns a stream with sensor information. This stream will feed the unmarshal() operation to convert the received sensors into a set of generic sensor objects defined in ViSIoT Core.

**Selector:** ranks the available sensors according to the best trade-off between their attributes. Different algorithms can be applied in this class to perform the suitable selection according to an specific condition.

Core: coordinates the communication with the other blocks. A GenericSensor component contains the at-tributes used to characterise the sensors that will be unmarshaled. This component can be extended to sup-port more attributes and functionalities according to each repository.

**Target :** marshals the GenericSensor component into the target format such as a file or another object. It must implement the marshal() operation to describe all sensor information used by the target environment to retrieve the data from the virtual sensor source.

## 4.6 Target system

The target system is composed by a set of target devices where the desired sensors will be deployed using a Vi-SIoT client application. In our prototype we set the GSN as our target system. The target system needs to able to deploy the virtual-sensor generated by ViSIoT and must have a wrapper capable of retrieving all information from the virtual-sensor. The target device will receive a RESTful message with the set of compressed files to be downloaded from the ViSIoT tool. After downloading the files are uncompressed and deployed into the target system, and an acknowledgement message reporting the success of the deployment is sent to the ViSIoT tool.

## 5. CASE STUDY

Rheumatism is a general term used to describe a group of diseases which affects joints, muscles and bones, characterized by pain and movement constraints. Common rheumatic disorders currently recognized includes osteoarthritis, rheumatoid arthritis, or fibromyalgia [Arnett et al., 1988].

Several works like Guedj and Weinberger (1990), Strusberg et al., (2002), Verges et al. (2004) correlates pain to changes in weather conditions such as temperature, air pressure and humidity. Our case study is based on Strusberg et al. (2002), which shows the relationship between weather and arthritis pain in 151 people with rheumatoid arthritis, osteoarthritis or fibromyalgia and a control group composed by 32 people without arthritis.

The results shows that patients in all three groups suffered significantly more pain on low temperature days. Also, the results showed that osteoarthritis patients were affected by high humidity, arthritis patients were affected by high humidity and high air pressure and fibromyalgia patients were affected by high air pressure [Strusberg et al., 2002].

Table 2 shows the correlation between the pain and the weather conditions.

Table II Correlation between Rheumatism diseases and weather conditions

| Weather Condition | Disease | | |
|---|---|---|---|
| | osteoarthritis | arthritis | fibromyalgia |
| Temperature | low | low | low |
| Humidity | high | high | Doesn't affect |
| Air Pressure | Doesn't affect | high | high |

Considering the winter season in the northern hemisphere (December 21 ~ March 20) and the existent correlation between rheumatism diseases and weather conditions, a doctor wants to compile a list to his patients about which cities had the best conditions in North America and Europe for people who suffers rheumatism diseases.

# 6. EXPERIMENT METHODOLOGY

We have used the ViSIoT to deploy the sensors in two GSN instances. Due to the high number of sensors available in each region (2862 for North America and 5184 for Europe) and infrastructure limitations, we split our experiments in two parts. In Section 6.1 we detail the performed experiments while in Section 6.2 we describe the environment specifications.

## 6.1 Experiment Design

In the first part (Section 7.1), we have limited the number of sensors used in our experiment to represent the main cities in each region as shown in Figure 3 for Europe and thus reduce the gathered data amount. Table III presents the latitude and longitude used to represent the regions in our experiments, which are visually represented in Figure 2. Then, a weather analysis is presented and correlated with rheumatism diseases presented in Table II. The interval between requests to the OpenWeatherMap is set to 30 minutes.

Table III. Qualitative experiment setup

| Region | Initial Point | | Final Point | | Available Sensors | Used Sensors |
|---|---|---|---|---|---|---|
| | Long. | Lat. | Long. | Lat. | | |
| North America | -170 | 70 | -60 | 30 | 2862 | 129 |
| Europe | -30 | 80 | 30 | 40 | 5184 | 130 |

The second part (Section 7.2) presents a performance evaluation of ViSIoT. In order to demonstrate the scalability of ViSIoT, experiments that combine the number of target devices (1, 4 and 16) and virtual sensors (1,000, 20,000, 40,000, 60,000, 80,000 and 100,000) were con-ducted to observe the time spent to setup a distributed environment.

Table 4 summarizes the experiment design used in the second part of experiments.

Table IV. Experiment Design

| Factor | Level |
|---|---|
| Number of Target Devices | 1, 4 and 16 |
| Number of Virtual Sensors | 1,000, 20,000, 40,000 60,000 , 80,000 and 100,000 |

We also assumed that the sensors retrieved from the repository had 6 context properties (battery, price, drift, frequency, energy consumption and response time) that were syntactically generated to allows the execution of the selector module. Each experiment was replicated 50 times to reduce the effects of stochastic components, which adds noisy to the gathered results.

The setup time is composed by the unmarshal, selection, marshal and deploy time. The time to request sensor meta-data is not considered because of their non-deterministic behavior. As depicted in Figure 1. ViSIoT Sequence Diagram, the un-marshal time represents how long it takes to unmarshal the message received to a generic object. The selection time corresponds the time spent to rank all available options. The marshal time express the time spent to generate the resources that will be used by the target devices. The deploy time corresponds to the sum of elapsed time to compress, transfer and uncompress the specified files to each target device, and returns an ACK to the user.

## 6.2 Environment Configuration

Our environment setup is composed of two kind of machines. A virtual node is used to host the ViSIoT prototype while physical machines are used to host the GSN server instances. Table V and Table VI shows the machine specifications of ViSIoT and the GSN server configura-tion. Both environments use Java 1.6 and Apache Tomcat 7.0.

Table V. ViSIoT Virtual node specification

| Hardware | Specification |
|---|---|
| Processor | 4 cores |
| Memory | 4 GB RAM |
| Motherboard | - |
| HD | 5GB |
| Operational System | Linux Ubuntu Server 14.04.1 LTS  64 Bits |
| Switch | Switch 3Com 2920-SFP Plus 16 ports Gigabit Switch  3CRBSG209 |

Table VI. GSN server specifications

| Hardware | Specification |
|---|---|
| Processor | Intel Core2 Quad |
| | Processor Q9400 |
| Memory | 8 GB RAM DDR3 Kingston |
| Motherboard | Gigabyte G41-MT-S2P |
| HD | 160GB Seagate |
| | Sata II 7200RPM |
| Operational System | Linux Ubuntu Server |
| | 14.04.1 LTS  64 Bits |
| Switch | Switch 3Com 2920-SFP Plus |
| | 16 ports Gigabit Switch  3CRBSG209 |

# 7. RESULTS

In this Section we present the experiment data and the performance results of the experiments described in Section 6. We describe the experiment data results in Section 7.1 and the performance evaluation results in Section 7.2.

## 7.1 Gathered Data Results

In this Section we present the gathered experiment data. Due to the huge number of available cities and sensors, we chose to analyse the extremes for temperature, air pressure and humidity. Table VII show the group of cities that presented the highest and lowest temperature (T(K)) with their air pressure (A.P(PA)) and humidity (H(%)) indexes in Europe and North America between February 7 and 9, 2015.

TABLE VII. HIGH AND LOW MEAN TEMPERATURES VALUES PRESENTED IN EUROPE AND NORTH AMERICA

| Region | W.C | City | Country | T(K) | A.P(PA) | H(%) |
|---|---|---|---|---|---|---|
| EU | High | Vlore | AL | 283.54 | 1015.59 | 100.00 |
| | | Bari | IT | 283.29 | 1022.26 | 100.00 |
| | | Gijon | ES | 282.52 | 1038.20 | 100.00 |
| | Low | Joensuu | FI | 266.78 | 998.36 | 80.62 |
| | | Kuusamo | FI | 265.65 | 974.30 | 82.79 |
| | | Longyearbyen | SJ | 246.35 | 966.72 | 60.51 |
| NA | High | Phoenix | US | 291.42 | 966.45 | 48.93 |
| | | Mexicali | MX | 290.61 | 1025.91 | 54.17 |
| | | Hamilton | BM | 290.34 | 1033.54 | 100.00 |
| | Low | Whitehorse | CA | 242.15 | 865.08 | 53.63 |
| | | Yellowknife | CA | 239.80 | 1011.24 | 52.47 |
| | | Fairbanks | US | 233.18 | 999.01 | 16.30 |

According to Table VII, Vlore (283.54 K) and Longyearbyen (246.35 K) represents the cities with the highest and lowest temperature in Europe, while Phoenix (291.42 K) and Fairbanks (233.18 K) showed the highest and lowest temperature in North America. Because of its larger size, North America has greater differences between the minimum and maximum indexes in temperature, pressure and humidity than those found in Europe. According to Table II, all the three types of rheumatism are sensitive to low temperatures. In this sense, it is extremely important for people who have rheumatism diseases to find the cities with the temperature index closer to the superior limits in each region.

Nevertheless, the temperature is only one factor that influence the pain in people with rheumatism diseases. Fibromyalgia patients also suffer influence from high levels of air pressure. Thus, Phoenix located in North America was the unique city, which presented suitable conditions for these kind of patients as presented high temperature (291.42K) and low air pressure (966.45PA). People who suffer osteoarthritis have their pain condition worsened when the humidity levels increase. Phoenix and Mexicali located in North America were the only cities which showed suitable conditions for these kind of patients as presented high temperature (291.42K and 290.61K) and low humidity levels (48.93% and 54.17%).

The last case handles arthritis patients, which have their pain condition worsened as the temperature decreases, and the humidity and pressure indexes increases. In this specific case, Phoenix located in North America was the unique city which presented the proper conditions for these kind of patients as it also presented suitable conditions for fibromyalgia and osteoarthritis patients. On the other hand, the other cities showed in Table VII located in both continent do not presented suitable conditions for any kind of patients as they do not agreed at least with one condition.

Under these circumstances, ViSIoT helps to setup a distributed environment using a graphical interface. The sensors deployed in the target environment enable to compile a list ranking the cities, which show the best weather conditions in North America and Europe for people who suffer from rheumatism diseases.

## 7.2 ViSIoT Performance Analysis

In this Section, we present the performance results of ViSIoT environment setup.

Figure 4a presents the time to unmarshal virtual sen-sors into generic objects. The time spent to unmarshal the messages slightly change according the number of sensors requested, and so this operation is scalable in the entire range (from 1,000 to 100,000).

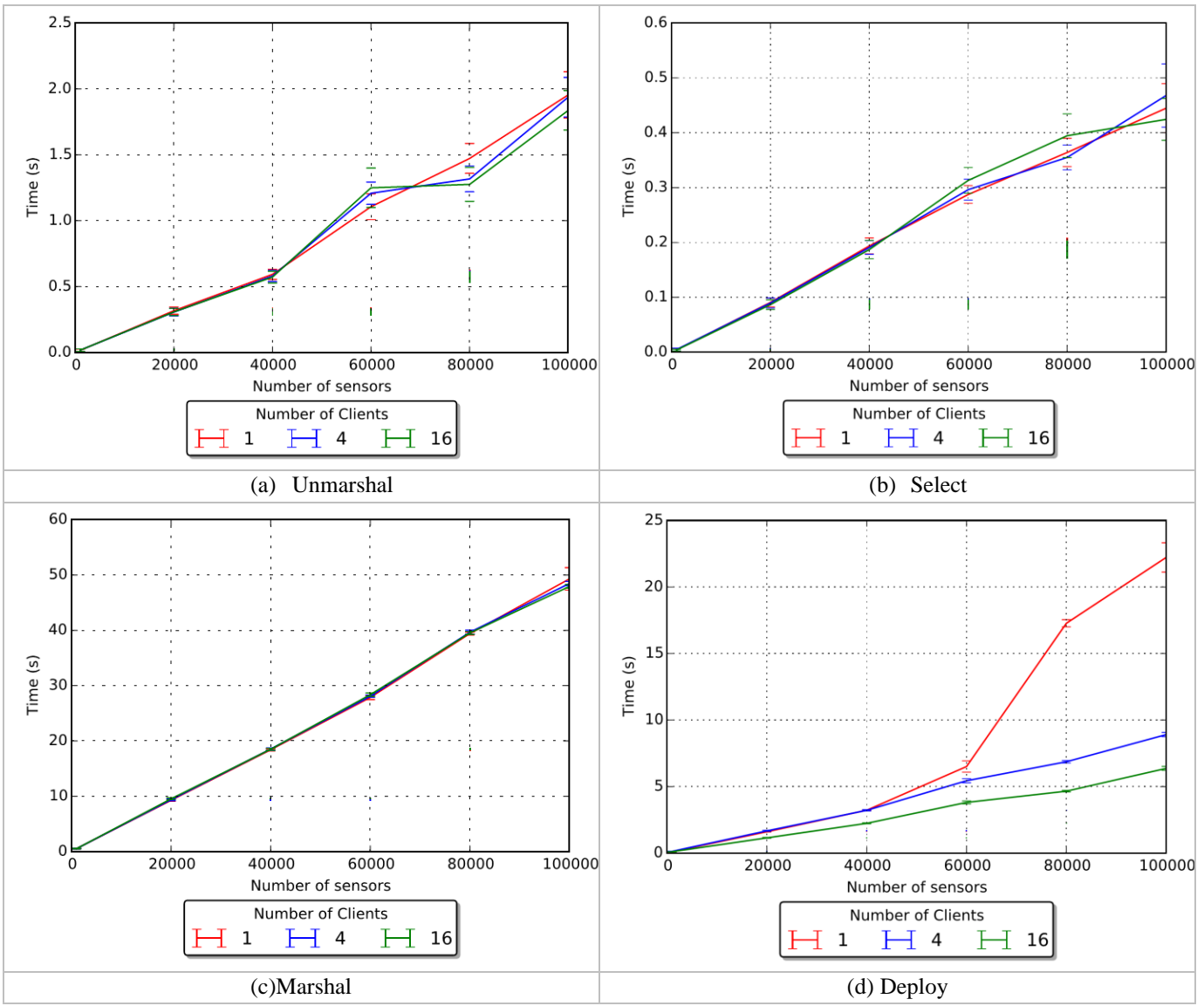

*Figure 4. Performance Results*

Figure 4b shows the times to rank the generic objects to an specific context using the TOPSIS algorithm. The time to rank the available sensors is extremely short as it uses the generic objects which are in the main memory of ViSIoT.

Figure 4c shows the times to marshal the generic objects into GSN virtual sensor files. The time spent to marshal is bigger than that observed in Figure 4a. This behaviour occurs because several operations of input and output are performed by the server side to generate the virtual sensor files.

Figure 4d presents the times to deploy the virtual sensor files to target devices. Similar to Figures 4a and 4c, the deploy time increases proportionally to the number of sensors to be deployed. However, it can be observed that the number of target devices influences the deploy time because the amount of objects to be transferred for each one decreases, which reduces the input/output operations that will be performed by them. Also, the compression mechanism aids to reduce the amount of time used to transfer the files, as only one request per client must be performed to ViSIoT.

It is important to highlight that the number of devices receiving the sensors descriptions does not influence in times for unmarshalling, selecting and marshalling, because these are general phases used to retrieve the sensors and generate the files that will be used without considering the number of target devices.

In summary, from the results, ViSIoT is scalable when considering a range of 1,000 to 100,000 virtual sensor to be deployed. The marshal operation is more time consuming than unmarshal and select operations. In the deploy time, the number of virtual sensors increases the absolute deploy time while the number of target devices decreases this time. The compressing mechanism improves the performance by shrinking the amount of data being transferred by the environment deployment. In addition, it is worth to mention that each of these operations consumes less than 1 minute. Hence, ViSIoT performed efficiently by spending less than 2 minutes for deploying virtual sensors, which enables dynamic con-figurations become feasible in distributed environments.

# 8. CONCLUSION

In this paper, we presented the ViSIoT platform which is a visual platform to provide sensors as a service in a global scale. ViSIoT is designed to support several independent sensor repositories and middleware for the Sensing as a Service. As a proof of concept, we built a working prototype to demonstrate the functionalities offered by ViSIoT. The ViSIoT performance analyses shows the capacity for setting up the environment in a timely manner. Also, a real use of ViSIoT is applied in the weather condition analyses to determine which cities present the better conditions to host people with different kinds of rheumatism diseases.

As future work, we intend to apply well-known sensor search and selection techniques such as presented in Elahi et al. (2009), Calbimonte et al. (2011) and Perera et al. (2013) in our tool to analyse the quality of the offered selection, and also develop our own search, selection and fault tolerance mechanisms. As smart spaces tend to be dynamics, ViSIoT can be evaluated considering an environment where changes are commonplace, as it can react in 2 minutes for an amount of virtual sensors in a magnitude of 100,000.

# 9. ACKNOWLEDGMENT

We thank Coordination of Improvement of Personal Higher Education (CAPES) and São Paulo Research Foundation (FAPESP, processes 2011/09524-7, 2013/26420-6 and 2011/12670-5), for the support of this research. We also thank ICMC-USP and LaSDPC for offering the necessary equipment for this study. Dr. Charith Perera's work is funded by European Research Council Advanced Grant 291652.

## Authors

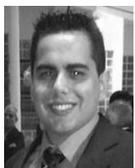

Graduated in Bachelor of Computer Science from Universidade Estadual Paulista Julio de Mesquita Filho (2011) and master's degree in Computer Science and Computational Mathematics from the University of S ão Paulo (2014). He is currently a doctoral student in the Institute of Computational Mathematics and Computer Science (ICMC) at University of S ão Paulo (USP). His main research topics are Internet of Things, Quality of Service, Service Level Agreement, Sensing the Service, Cloud Computing and Wireless Sensor Networks.

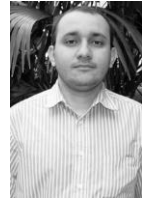

J úlio C. Estrella received the B.Sc in computer science Universidade Estadual Paulista Julio de Mesquita Filho in 2002 and M.Sc. and Ph.D. degrees in computer science from the University of Sao Paulo (USP), Sao Paulo, Brazil, , 2006, and 2010, respectively. He has experience in Computer Science with emphasis in Computer Systems Architecture, acting on the following themes: service oriented architectures, web services, performance evaluation, distributed systems, computer networks and computer security. He is currently an Assistant Professor with the Institute of Mathematics and Computer Science (ICMC), USP.

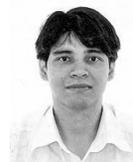

Luis H. V. Nakamura is a PhD candidate in Institute of Computer Science and Mathematics Computational (ICMC) at University of S ão Paulo (USP). He received a B.S from Technology College (FATEC) in 2006, and a M.S. from the University of S ão Paulo (ICMC-USP) in 2012. His research interests are based on distributed systems, which includes Cloud Computing, Autonomic Computing and Semantic Web. He has also investigated the implications of Web Services and Performance Evaluation of computational systems.

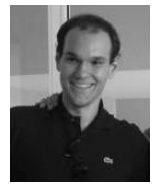

Rafael M. de O. Libardi holds a BA in Informatics at Computer Science and Computational Mathematics Institute - USP in S ão Carlos (2013) and is currently doing Masters also at ICMC-USP addressing security in cloud environments.

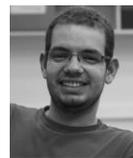

Carlos H. G. Ferreira Graduated in Information Systems from the Federal University of Vi çosa - Campus Rio Parana ́ba. Currently he is a Master Degree candidate in Computer Science at Institute of Mathematical Sciences and Computing, University of S ão Paulo acting in the following lines: Service Oriented Architectures, Web Services, Cloud Computing, and Computer System Performance Evaluation.

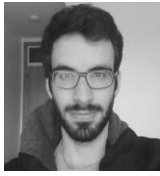

Liuri Loami is a Computer Science student at University of Sao Paulo, with experience with robotics, web development and project management. He is currently enrolled at Dublin Institute of Technology as an international student, supported by the Science without Borders program.

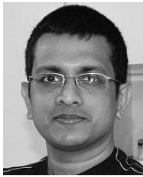

Charith Perera is a Research Associate at The Open University, UK. Currently, he is working on the Adaptive Security and Privacy (ASAP) research programme. He received his BSc (Hons) in Computer Science in 2009 from Staffordshire University, Stoke-on-Trent, United Kingdom and MBA in Business Administration in 2012 from University of Wales, Cardiff, United Kingdom and PhD in Computer Science at The Australian National University, Canberra, Australia. Previously, he worked at Information Engineering Laboratory, ICT Centre, CSIRO. His research interests are Internet of Things, Sensing as a Service, Privacy, Middleware Platforms, Sensing Infrastructure. He is a member of both IEEE and ACM. Contact him at charith.perera@ieee.org .

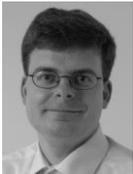

Stephan Reiff-Marganiec is a Senior Lecturer in Computer Science at the University of Leicester. He has worked in the computer industry in Germany and Luxembourg and held research positions at the University of Glasgow (while simultaneously reading for a PhD) and the University of th and 10th International Conference on Feature Interactions in Telecommunications and Software Systems an was co-Chair of three instances of YR-SOC. Stephan lead workpackages in the EU funded projects Leg2Net, Sensoria and inContext focusing on automatic service adaption, context aware service selection, workflows and rule based service composition. Stephan is co-editor of the Handbook of Research on Service-Oriented Systems and Non-Functional Properties and has published in excess of 50 papers in international conferences and journals as well as having served on a large number of programme committees. Stephan was appointed Guest Professor at the China University of Petroleum and was visiting Professor at Lamsade at the University of Dauphine, Paris. He was elected Fellow of the BCS (FBCS) in 2009 and is a member in both ACM and IEEE.